\documentclass{midl} 

\usepackage{xcolor}
\usepackage{graphicx}
\graphicspath{{figures/}}
\usepackage[font=small,labelfont=bf]{caption}

\jmlrproceedings{MIDL}{Medical Imaging with Deep Learning}
\jmlrpages{}
\jmlryear{2019}
\jmlrworkshop{MIDL 2019 -- Extended Abstract Track}

\title{GANs 'N Lungs: improving pneumonia prediction}

\midlauthor{\Name{Tatiana Malygina\nametag{$^{1,2}$}} \Email{tanya.malygina@botkin.ai}\\
\addr $^{1}$ Intellogic Limited Liability Company (Intellogic LLC), office 1/334/63, building 1, 42 Bolshoi blvd., territory of Skolkovo Innovation Center, 121205, Moscow, Russia\\
\addr $^{2}$ The laboratory of bioinformatics, ITMO University, Kronverkskiy pr. 49, St. Petersburg, 197101, Russia \\
\AND
\Name{Elena Ericheva\midlotherjointauthor\nametag{$^{1}$}} \Email{elena.ericheva@botkin.ai}
\AND
\Name{Ivan Drokin\midlotherjointauthor\nametag{$^{1}$}} \Email{ivan.drokin@botkin.ai}
}

\setcitestyle{numbers,square}

\begin{document}

\maketitle

\begin{abstract}
We propose a novel method to improve deep learning model performance on highly-imbalanced tasks. The proposed method is based on CycleGAN to achieve balanced dataset. We show that data augmentation with GAN helps to improve accuracy of pneumonia binary classification task even if the generative network was trained on the same training dataset.
\end{abstract}


\section{Introduction}

The ability of generative adversarial networks (GAN) to generate images made them a useful tool in many research areas and in generative art. It was shown~\cite{DBLP:journals/corr/abs-1803-01229} that GANs can also be used to improve model training by extending train dataset with generated, previously unseen images.

One of significant problems and limitations of automatic medical image processing is the fact that the numbers of target lesions of different types are highly unbalanced. Many irregular lesions that are visible on medical image do not belong to the target class. As well as number of samples from normal class is overwhelming \cite{cxr_dataset}. 
As the result, the extraction of main characteristics of lesions is a complex process and requires many heuristic steps. Preprocessing that brings all objects and visualization methods to the same distribution is still an important step that affects the efficiency of the classification. 
We have designed system for dataset balancing via oversampling with CycleGAN, that deals well with substantially unbalanced datasets.



\section{Data augmentation framework} 


We consider binary classification problem. We start with dataset $X$ consisting of $n$ samples $x_1, x_2, \dots, x_n$ and corresponding labels $l_1, l_2,\dots,l_n$, where $l_i\in\{0,1\}$ for $i=1,\dots,n$. Suppose that $X$ is imbalanced, i.e. one class of labels dominates over another. To deal with this problem, we propose oversampling using generative adversarial network (GAN). We employ two networks $G_{0\rightarrow1}$ and $G_{1\rightarrow0}$, that produce complementary samples. E.g., for sample $x_i$ with a corresponding label $l_i=0$ a new sample $\tilde{x}_i=G_{0\rightarrow1}(x_i)$ with $\tilde{l}_i=1$ is generated, and for sample $x_j$ with $l_j=1$ a sample $\tilde{x}_j=G_{1\rightarrow0}(x_j)$ with $\tilde{l}_j=0$ is generated. These $G_{0\rightarrow1}$ and $G_{1\rightarrow0}$ models can be constructed separately with any image-to-image generative networks.

We combine newly generated samples $\tilde{X}$ with samples from $X$ to produce perfectly balanced dataset $X_{\mbox{aug}}$ consisting of samples $x_1,\dots,x_n,\tilde{x}_1,\dots,\tilde{x}_n$, with labels  $l_1,\dots,l_n,\tilde{l}_1=1-l_1,\dots,\tilde{l}_n=1-l_n$. We use $\tilde{X}$ to train binary classifiers.



\section{Experiments}

We conducted our experiments on imbalanced dataset with X-ray images subsampled from CheXnext~\cite{cheXnext} dataset with preserved original train/validation split. We've selected images with pneumonia (with label 1 in Pneumonia class) or with no abnormalities (with $l_i = 0$ for all $i$).

\begin{figure}
\centering
\begin{minipage}{.48\textwidth}
  \centering
  \includegraphics[width=.95\linewidth]{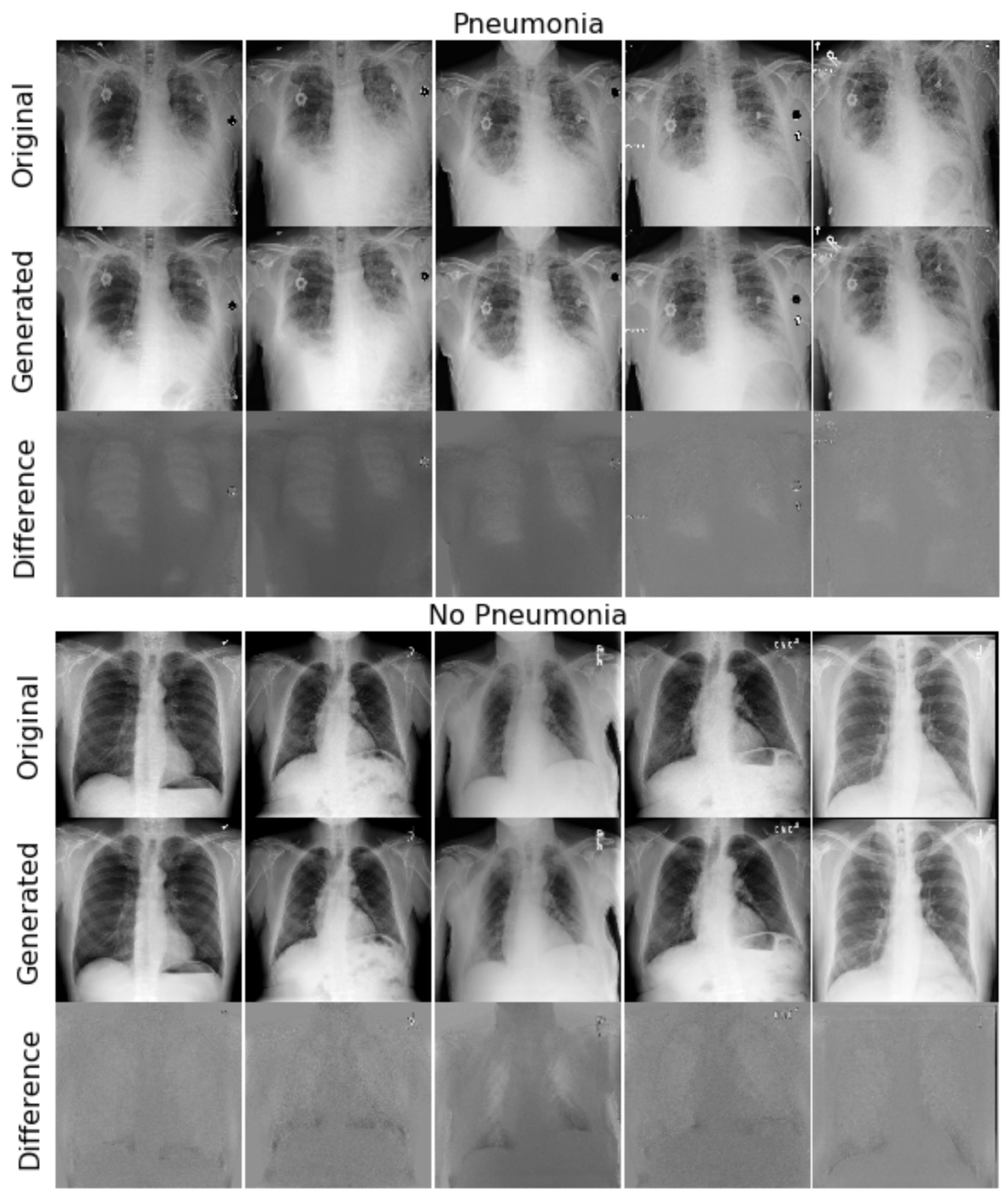}
	\captionof{figure}{Picture shows Original images from CheXnext and their CycleGAN-generated pairs. Difference between Original and Generated images is shown in Difference row.}
	\label{fig1}
\end{minipage}%
\hspace{0.2cm}
\begin{minipage}{.48\textwidth}
  \centering
  \includegraphics[width=.95\linewidth]{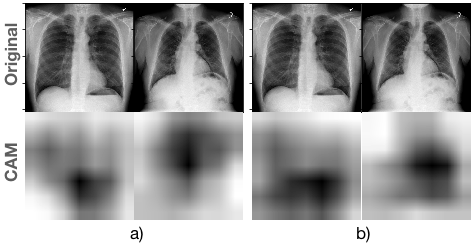}
  \captionof{figure}{Class activation maps example from models: a) trained without augmentation; b) trained with augmentation.}
  \label{fig:images_with_CAM}
  \includegraphics[width=.95\linewidth]{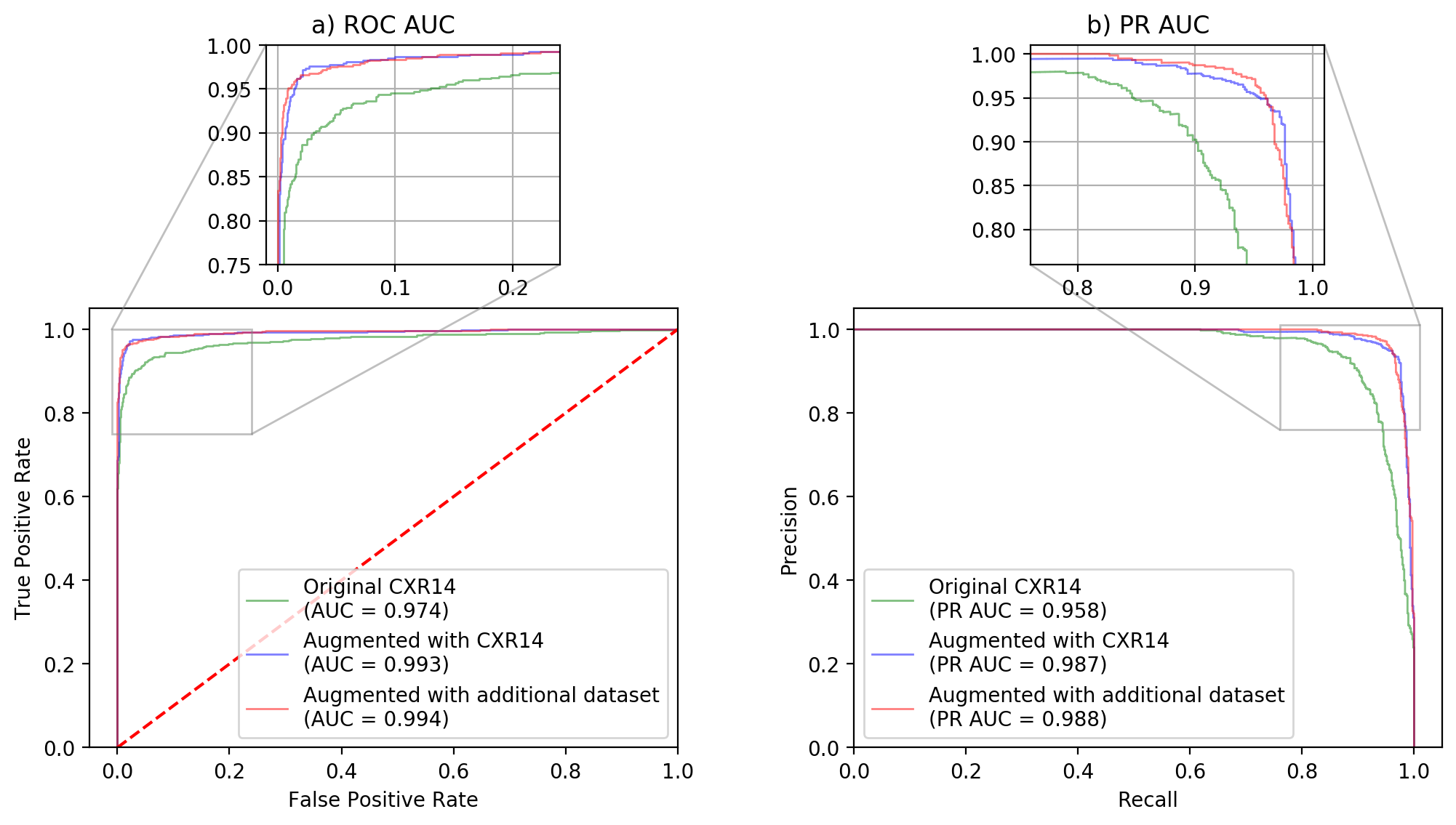}
  \captionof{figure}{Test metrics for pure CheXnext subset and GAN-augmented datasets: a) ROC curve; b) Precision-Recall curve.}
  \label{fig:roc_prcurve}
\end{minipage}
\end{figure}

CycleGAN~\cite{CycleGAN2017} with cycle-consistency loss was used to construct $G_{0\rightarrow1}$ and $G_{1\rightarrow0}$ generators.  We've used them to produce X-ray images with pneumonia from images with no pneumonia, and vice versa. Figure~\ref{fig1} shows several pairs of prototype images from CheXnext, corresponding complementary images generated with CycleGAN and the difference between these images.


Our binary classifier is based on DenseNet-121~\cite{DenseNet} architecture -- in this way it is similar to the CheXnet and CheXnext models. It expects $224\times 224$ images as the input and produces probabilities for target classes. As we focus on solving binary classification task, we use binary cross-entropy loss to optimize model during training. 



To compare DenseNet model performance with and without augmentation, we use the same validation dataset. We compare results for 3 train datasets in binary classification task: pure CXR14 (no augmentation was used); augmented with CycleGAN pretrained on CXR14 dataset; augmented with CycleGAN, which was pretrained on additional dataset~\cite{kaggle_dataset} . ROC and precision-recall curves for all classifiers are shown at Figure~\ref{fig:roc_prcurve}.
You can see that with our balancing method ROC AUC score increased from 0.9745 to 0.9929 (0.9939 if we use additional dataset to train CycleGAN), PR AUC value increased from 0.9580 to .9865 (0.9883 if we use additional dataset to train CycleGAN) on validation data.


Also we provide examples of class activation maps~\cite{zhou2016cvpr} on Figure~\ref{fig:images_with_CAM}. It highlights the most informative image regions relevant to the predicted class. 





\section{Conclusion and Future Work}

The proposed data augmentation method has improved performance of binary classifier in biomedical field, and it might be useful in other computer vision tasks.


Despite CheXnext dataset is quite large even for binary classification task and all experiments are conducted with very limited hardware resources (single NVIDIA 1070 graphical card) the method produce visually-distinguishable classes of image. Thus this augmentation technique does not require extensive computational resources and could be applied in everyday machine learning practice.

For non-radiologist it might be unclear if generated images belong to pneumonia or no pneumonia class (see Figure~\ref{fig1}). 
That's why we plan to test both the model and augmented images against human radiologists on different datasets, including private datasets.




We plan to develop effective procedure and extend proposed augmentation method for multi-class tasks as well as for multi-label datasets, i.e. full CheXnext dataset with 14 classes, in cross-domain CycleGAN for x-rays/fluorography.We also plan to run a more detailed analysis of $G_{0\rightarrow1}$ and $G_{1\rightarrow0}$ models and explore classification model robustness against generative model output.


\bibliography{malygina19}
\end{document}